\documentstyle[aps]{revtex}

\begin{document}

\begin{flushright}
e-print JLAB-THY-97-25 \\
June 1997 \\
%hep-ph/
\end{flushright}

\vspace{2cm}

\begin{center}
{\Large \bf Asymmetric Parton 
Distributions}
\end{center}
\begin{center}
{\sc Anatoly V. Radyushkin\footnote{Also at Laboratory of Theoretical Physics,
JINR, Dubna, Russia}} \\ 
{\em Physics Department, Old Dominion University,
 Norfolk, VA 23529, USA}  \\ {\em
 Jefferson Lab, Newport News,VA 23606, USA}

\end{center}
\vspace{2cm}

\begin{abstract}
Applications of perturbative QCD to
hard exclusive electroproduction 
processes in the Bjorken limit at 
 small invariant momentum transfer $t$ 
bring in a  new type of 
parton distributions
which have 
hybrid properties, 
 resembling both the parton distribution functions$f_a(x)$
and  the distribution
amplitudes. Their $t$-dependence
is analogous to that of  hadronic form factors.  
We discuss general properties of these new parton distributions, 
their relation to 
usual parton densities 
and the evolution equations which they satisfy.
\end{abstract}

\section*{Introduction} 

The use of  phenomenological
functions  accumulating information about
nonperturbative long-distance dynamics
is the standard feature of  
applications of pQCD to hard
processes. 
The well-known 
 examples are the 
 parton distribution functions $f_{p/H}(x)$ \cite{feynman}
used in  pQCD approaches to hard
inclusive processes and 
distribution amplitudes   $\varphi_{\pi}(x)$ \cite{tmf} {\it etc.}
which naturally
emerge in the  asymptotic  QCD
analyses of hard exclusive processes. 
Recently,  it was argued that the same 
gluon density  $f_g(x)$ used for description
of hard inclusive processes also determines 
the amplitudes  of  hard exclusive $J/ \psi$ \cite{ryskin}
and  $\rho$-meson 
\cite{bfgms} electroproduction. 
It was also proposed \cite{ji} 
to use  another exclusive process of  
deeply virtual Compton scattering $\gamma^* p \to \gamma p' $
 (DVCS) for measuring   
parton  distribution functions inaccessible 
in  inclusive measurements. An  important feature 
(noticed  long ago  \cite{barloe,glr,drm})
 is that  kinematics of DVCS and hard elastic 
electroproduction processes, in a frame where the hadron moves fast,
requires 
the presence of the longitudinal component
in the  momentum transfer 
$r\equiv p - p'$ from the initial hadron to the final:
$r_{\|} = \zeta p$.
This means that the matrix element
$\langle p' | \ldots | p \rangle$ 
is essentially asymmetric (``skewed'').
For DVCS and $\rho$-electroproduction 
in the region       $Q^2 >> |t|, m_H^2$,
the ``skewedness''  parameter $\zeta$
coincides with  the
Bjorken variable $x_{Bj} = Q^2/2(pq)$.
For diffractive processes, 
both the  skewedness  $\zeta$ 
of the nonperturbative
matrix element $\langle p' | \ldots | p \rangle$ 
and the absolute value of the momentum transfer $t \equiv (p'-p)^2$
are small.  Studying  the DVCS process,
one should be able to  consider the whole region 
\mbox{$0 \leq \zeta \leq 1$}  and $t \sim  1\, GeV^2$ \cite{ji2}. 
In this situation, one deals with essentially
{\it nonforward} (or {\it off-forward} in terminology 
of ref. \cite{ji})
matrix element $\langle p' | \ldots | p \rangle$. 
The   pQCD approaches   
incorporating  asymmetric/off-forward  parton distributions 
were formulated in refs.\cite{ji,ji2,compton,gluon,evol}.
A detailed analysis of pQCD factorization 
for hard meson electroproduction processes
was given in ref.\cite{cfs}.
The basic idea of our papers \cite{compton,gluon}  is that 
one should treat the initial momentum $p$ and 
the longitudinal part of the momentum transfer $r$ on 
equal footing by  introducing 
double distributions
$F(x,y)$, which specify the fractions of  $p$ 
and  $r_{\|}$ 
carried by the active parton: $k_{\|} = xp + y r_{\|}$. 
These distributions look like distribution functions 
with respect to $x$ and like  distribution amplitudes
with respect to $y$.
The  longitudinal momentum transfer  
$r_{\|}$ is proportional to $p$: $r_{\|} = \zeta p$
and  it is  convenient to parametrize
matrix elements $\langle p-r | \ldots  |p \rangle$
by  {\it nonforward } ${\cal F}_{\zeta} (X;t)$ or, in the $t \to 0$ limit, 
 {\it asymmetric distribution functions}
${\cal F}_{\zeta} (X)$  specifying the total light-cone fractions
$Xp=(x+y \zeta )p$, $(X-\zeta)p\equiv X'p$   
of  the initial hadron momentum $p$
carried by the ``outgoing'' and ``returning'' 
partons\footnote{Asymmetric distribution functions 
defined in ref.\cite{gluon} 
are  similar  to, but not  coinciding  with
the $t \to 0$ limit of the
off-forward parton distributions 
introduced by X.Ji \cite{ji}.}.  
Double 
distributions $F(x,y)$ are universal functions in the sense that
they do not depend on the skewedness 
parameter $\zeta$ while the asymmetric  distribution
functions ${\cal F}_{\zeta} (X)$  form a family 
of $X$-dependent functions changing their shape when 
$\zeta$ is changed\footnote{A more detailed
discussion of double distributions and their relation to
nonforward  distributions was given in my talk about DVCS \cite{dvcs}.}.
The functions ${\cal F}_{\zeta} (X)$ 
also have hybrid properties. 
When $X \geq \zeta$,  the
returning parton   carries a positive
fraction $(X-\zeta)p$ of the initial momentum $p$,
and  ${\cal F}_{\zeta} (X)$  is similar 
to the usual parton distribution $f(X)$.
In the region $0 \leq X \leq \zeta$
the difference $X- \zeta$ is  negative,
$i.e.,$ the second parton should be  treated
as propagating together with the first one.
The partons in this case share 
the longitudinal momentum transfer $r_{\|} =\zeta p$ 
in fractions $Y\equiv X/\zeta$ and $1-Y$: 
in the region $X \leq \zeta$
the function ${\cal F}_{\zeta} (X)$ 
looks like a distribution amplitude.

It is very important to know  spectral 
properties of the  parton distributions $F(x,y)$
and ${\cal F}_{\zeta} (X)$. 
In ref. \cite{evol}, it was shown 
that  both arguments of a double distribution
$F(x,y)$ 
satisfy the natural ``parton'' constraints 
 $0 \leq x \leq 1$, $ 0 \leq y \leq 1$
for  any Feynman  diagram contributing  to $F(x,y)$. 
A less obvious restriction   $0 \leq x+y \leq 1$ 
 \cite{compton,gluon,evol} guarantees 
that the argument $X=x+y \zeta$ of the  asymmetric distribution
${\cal F}_{\zeta} (X)$ also changes between 
$0$ and $1$. Since $X=0$ can be obtained only if both $x=0$
and $y=0$, due to   vanishing 
phase space for such a configuration
the asymmetric distributions ${\cal F}_{\zeta} (X)$ 
vanish for $X=0$.  This property is very essential,
because the hard subprocess amplitudes usually
contain $1/X$ factors.  When   ${\cal F}_{\zeta} (0) \neq 0$,
one faces a non-integrable singularity   ${\cal F}_{\zeta} (X)/X$  
at $X=0$.

\section*{Asymmetric  distributions}

There are two leading-twist quark  operators
 $\bar \psi_a (0) \gamma_{\mu} E(0,z;A)  \psi_a (z)$ and 
 $\bar \psi_a (0) \gamma_5 \gamma_{\mu} E(0,z;A)  \psi_a (z)$, where 
$E(0,z;A)$ is the usual path-ordered exponential 
which makes the operators gauge-invariant. 
 In the forward case, 
the first one gives the  spin-averaged densities  $f_a(x)$
while the second one is related to the spin-dependent ones 
$\Delta f_a(x)$. Here, we
will discuss the 
$\bar \psi_a  \gamma_{\mu} E(0,z;A)  \psi_a$
operators and gluonic operators with
which they  mix under evolution.  
The relevant nonforward matrix element 
can be written as
\begin{eqnarray} 
\hspace{-1cm} 
&&\lefteqn{
\langle \, p'  , s' \,  | \,   \bar \psi_a(0)  \hat z 
E(0,z;A)  \psi_a(z) 
\, | \,p ,  s \,  \rangle |_{z^2=0}
 } \label{57}  \\ \hspace{-1cm} &&  
= \bar u(p',  s')  \hat z 
 u(p,s) \, \int_0^1   
 \left ( e^{-iX(pz)}
 {\cal F}^a_{\zeta}(X;t)  -  e^{i(X-\zeta)(pz)}
{\cal F}^{\bar a}_{\zeta}(X;t)  \right ) 
 \,  dX \nonumber \\ \hspace{-1cm} &&  
+
\, \bar u(p',s')  \frac {\hat z  \hat r - \hat r \hat z}{2M}
 u(p,s) \, \int_0^1   
  \left ( e^{-iX(pz)}
 {\cal K}^a_{\zeta}(X;t)  -  e^{i(X-\zeta)(pz)}
{\cal K}^{\bar a}_{\zeta}(X;t)  \right ) 
 \,  dX
,
\nonumber
 \end{eqnarray} 
where  $\hat z \equiv z^{\mu} 
\gamma_{\mu}$. 
In Eq.(\ref{57}), quark and  antiquark contributions
are  explicitly separated (cf. \cite{lnc}). 
As emphasized by X. Ji \cite{ji}, 
the parametrization of this nonforward matrix element
must  include both the nonflip
 functions 
$ {\cal F}_{\zeta}(X;t)$
and the spin-flip  functions which we denote
by  ${\cal K}_{\zeta}(X;t)$. 
Taking the $O(z)$ term of the Taylor expansion gives the 
sum rules \cite{ji} 
\begin{eqnarray} 
\int_0^1   
 \left [
 {\cal F}^a_{\zeta}(X;t)  -  
{\cal F}^{\bar a}_{\zeta}(X;t)  \right ]
 \,  dX  =F^a_1(t) \, , \label{58}
\\
\int_0^1   
 \left [
 {\cal K}^a_{\zeta}(X;t)  -  
{\cal K}^{\bar a}_{\zeta}(X;t)  \right ] 
 \,  dX  =F^a_2(t)
\label{59} \end{eqnarray}
relating the nonforward distributions $ {\cal F}^a_{\zeta}(X;t)$,  
$ {\cal K}^a_{\zeta}(X;t)$ to 
the $a$-flavor components of the Dirac and Pauli
form factors, respectively.
For gluons, the    nonforward  distribution 
can be defined  
through the matrix element
\begin{eqnarray} 
&&
\langle p'  \,  | \,   
z_{\mu}  z_{\nu} G_{\mu \alpha}^a (0) E^{ab}(0,z;A) 
G_{ \alpha \nu }^b (z)\, | \,p  \rangle |_{z^2=0}
  \label{63}  \\  &&  
= \bar u(p')  \hat z 
 u(p) \, (z \cdot p) \, \int_0^1   
\frac1{2}  \left [ e^{-iX(pz)}
 + e^{i(X-\zeta)(pz)} \right ] 
 {\cal F}^g_{\zeta}(X;t) 
 \,  dX  + ``{\cal K}^g_{\zeta}" \,  .
 \end{eqnarray} 
In the formal $t=0$ limit, the nonforward distributions 
${\cal F}_{\zeta}(X;t)$, ${\cal K}_{\zeta}(X;t)$
convert into the asymmetric distribution functions
${\cal F}_{\zeta}(X)$, ${\cal K}_{\zeta}(X)$.
In the $\zeta =0$ limit, ${\cal F}_{\zeta}(X)$
reduce to the usual parton densities
$ {\cal F}^a_{\zeta=0}(X) = f_a(X) \, ;  \, 
{\cal F}^g_{\zeta=0}(X) = Xf_g(X).
$
For small $\zeta$, ${\cal F}^a_{\zeta}(X)$
differs from $f_a(X)$ by $O(\zeta)$.

\section*{Evolution equations}

Near  the light cone $z^2 \sim 0$, the bilocal operators 
$\bar \psi(0) \ldots \psi(z)$ 
develop logarithmic singularities $\ln z^2$, and  the 
limit $z^2 \to 0$ is singular.  Calculationally, these
singularities manifest themselves as  ultraviolet divergences
for the light-cone operators. 
The divergences are removed by  a 
subtraction  prescription characterized 
by some  scale $\mu$: 
${\cal F}_{\zeta} (X;t) \to {\cal F}_{\zeta} (X;t;\mu)$.
In QCD,  the gluonic 
operator
$
{\cal O}_g(uz,vz) =
z_{\mu}  z_{\nu} 
G_{\mu \alpha}^a (uz) E^{ab}(uz,vz;A) 
G_{ \alpha \nu}^b (vz) 
$
 mixes under renormalization with the flavor-singlet 
quark operator.
In the leading logarithm 
approximation, the easiest way to get the evolution equations 
for  nonforward distributions  is 
to use 
evolution equations \cite{brschweig,bb}
for  light-ray 
operators:
\begin{equation}
 \mu \, \frac{d}{d \mu} \,  
{\cal O}_a(0,z)    =
\int_0^1  \int_0^{1}  
\sum_{b} B_{ab}(u,v ) {\cal O}_b( uz, \bar vz) \,
\theta (u+v \leq 1) \, du \, d v  \,  , 
\label{74} \end{equation}
where $\bar v \equiv 1-v$ and $a,b = g,Q$. For valence distributions,
there is no mixing, and their  
 evolution 
is generated  by the $QQ$-kernel  alone.
Inserting Eq.(\ref{74})  between chosen  hadronic states
and parametrizing  matrix elements by appropriate
distributions, 
one can get DGLAP and BL-type   
kernels and also to calculate  the 
new  kernels  $R^{ab}(x,y;\xi,\eta)$ and  $W_{\zeta}^{ab}(X,Z)$.
The kernels $ R^{ab}(x,y;\xi,\eta)$
  govern the evolution of  
double  distributions:
\begin{equation} 
 \mu \frac{d}{d\mu}  \, F^a(x,y;t;\mu) =
\int_0^1 d \xi \int_0^{1-\xi} \, \sum_b \,  R^{ab}(x,y;\xi,\eta) \, 
  F^b(\xi,\eta;t;\mu) \, d \eta \, 
 .
\label{75} \end{equation}
The form of  equations is not affected by the
$t$-dependence,  and ``$t$''  will not be 
indicated in what follows.
Since  integration over $y$ converts $F^a(x,y)$ 
into the parton density  $f_a(x)$,
whose evolution is governed by the DGLAP equation,
the kernel $R_{QQ}(x,y; \xi, \eta)$ has  the property
\begin{equation}
\int_0^ {1-x}  R_{QQ}(x,y; \xi, \eta) d y =
\frac1{\xi} P_{QQ}(x/\xi).
\label{eq:rtop}
\label{114} \end{equation}
Integrating $R_{QQ}(x,y; \xi, \eta)$
over $x$ one  gets  the BL-type   kernel:
\begin{equation}
\int_0^{1-y}   R_{QQ}(x,y; \xi, \eta) d x = V_{QQ}(y,\eta).
\label{115} \end{equation}
Another set of kernels  $W_{\zeta}^{ab}(X,Z)$ 
describes the evolution of nonforward distributions 
and asymmetric distribution functions:
\begin{equation}
 \mu \frac{d}{d\mu}  {\cal F}_{\zeta}^a(X;t;\mu) =
\int_0^1  \, \sum_b \,  W_{\zeta}^{ab}(X,Z) \, 
{\cal F}_{\zeta}^b( Z;t;\mu) \, d Z \,  .
\label{76} 
 \end{equation}
When  $\zeta =1$,  the initial momentum 
coincides with the momentum transfer and  
${\cal F}_{\zeta}(X)$ reduces to a distribution amplitude
whose  evolution is governed by the 
BL-type  kernels:
$
W_{\zeta =1}^{ab}(X,Z)= V^{ab}(X,Z). 
$
The BL-type kernels  
appear as a  part of the asymmetric kernel $W_{\zeta }^{ab}(X,Z)$
even in the general $\zeta \neq 1,0$ case \cite{evol}.
As explained earlier, if   $X$  is  in the
region $X \leq \zeta$, 
then the  function ${\cal F}_{\zeta}(X)$ 
can  be treated as a distribution amplitude
$\Psi_{\zeta}(Y)$ with $Y= X/  \zeta$. 
For this reason, when both $X$ and $Z$ are smaller than $\zeta$,
the kernels 
$W_{\zeta}^{ab}(X,Z)$ simply  reduce 
to the  BL-type  evolution kernels $V^{ab}(X/\zeta,Z/\zeta)$.
 Furthermore, the BL-type kernels  also 
govern the evolution  in the region corresponding to transitions
from a fraction $Z$ which is larger
than $\zeta$ to a fraction $X$ which is smaller 
than $\zeta$. Note, that  the evolution jump through  the critical
fraction $\zeta$ is irreversible: 
$X \leq \zeta$ if  $Z \leq \zeta$. 
To put it in  words, when the parton momentum 
degrades in the evolution 
process to values smaller than the momentum transfer
$\zeta p \equiv r$,
further  evolution is like that for a distribution
amplitude: the momentum can decrease or increase
up to the $r$-value but cannot exceed this value.

When  $X > \zeta$,  
the initial quark momentum $Xp$ 
is larger than the momentum transfer $r = \zeta p$,
and the asymmetric distribution
 function ${\cal F}_{\zeta}^a  (X)$ 
is   a generalization of the 
usual  distribution function $f_a(X)$ for a 
 skewed kinematics. 
Hence, the evolution in the region 
$\zeta < X \leq 1$   ,  
 $\zeta < Z \leq 1$ is similar
to that generated by the  DGLAP 
equation.
 The relevant kernels $ P_{\zeta }^{ab}(X,Z)$
can be obtained from the light-ray 
kernels, $e.g.,$  
\begin{equation} 
\displaystyle  
P_{\zeta }^{QQ}(X,Z) 
= \frac{Z-X}{ZZ'} \int_0^{1}    \, 
B_{ab} \left   ( \bar w \, (1- X/Z)  \, , \, 
w\, (1- X'/Z')  \right ) dw \, , 
\label{94} \end{equation}
where $Z' \equiv Z - \zeta$.
  Note, that 
$P_{\zeta }^{QQ}(X,Z)$ is given by a function symmetric 
with respect to the interchange  of $X,Z$ with  $X',Z'$. 
In the $\zeta \to 0 $ limit, $P_{\zeta }^{QQ}(X,Z)$ 
converts into the DGLAP kernel:  $P_{\zeta =0}^{QQ}(X,Z) = P^{QQ}(X/Z)/Z$. 
 
{\it Acknowledgement.} This work was supported 
by the US Department of Energy
 under contract 
DE-AC05-84ER40150.

\end{document}